\documentclass[sort&compress,final]{aipproc}

\layoutstyle{6x9}

\begin{document}

\title{Highly anisotropic dissipative hydrodynamics}

\author{Michael Strickland}{
  address={Physics Department, Gettysburg College, Gettysburg, PA 17325 United States}
}



\begin{abstract}
The quark gluon plasma generated in ultrarelativistic heavy ion collisions may possess sizable
momentum-space anisotropies that cause the longitudinal and transverse pressures in the local rest
frame to be significantly different.  We review recent attempts to derive a dynamical framework that can
reliably describe systems that possess a high degree of momentum-space anisotropy.  The dynamical 
framework that has been developed can describe the evolution of the quark gluon plasma ranging 
from the longitudinal free-streaming limit to the ideal hydrodynamical limit.
\end{abstract}

\maketitle

\section{Introduction}

The Relativistic Heavy Ion Collider (RHIC) and the Large Hadron Collider (LHC) are studying the behavior of 
nuclear matter at high energy densities, $\epsilon \gg 1\;{\rm GeV/fm}^3$, using relativistic heavy ion 
collisions.  The goal of these experiments is not only to generate a deconfined quark-gluon plasma (QGP), 
but to also study its properties such as transport properties, color opacity, etc.  One complicating factor is 
that the QGP generated in such collisions lasts for only a few fm/c and during this time the bulk properties 
of the system, e.g. energy density and pressure, can change rapidly.  In addition, it is possible that in certain
regimes the system is far from equilibrium/isotropy.  Dynamical models that can reliably
describe the evolution of the system on the fm/c timescale are necessary in order to make sound 
phenomenological predictions.

One of the key outstanding questions in dynamical models of the QGP is to what extent is the QGP isotropic 
in the local rest frame.  Early indications from ideal hydrodynamical fits to experimental data for the elliptic 
flow of hadrons indicated that the data were consistent with isotropization at $\tau \sim 0.5$~fm/c after 
the initial nuclear impact \cite{Huovinen:2001cy,Hirano:2002ds}.  This rather short time 
scale prompted a plethora of papers attempting to solve the early isotropization/thermalization puzzle.  In 
the intervening ten years ideal hydrodynamics has been replaced by viscous hydrodynamics 
\cite{Israel:1979wp,Muronga:2006zx,Luzum:2008cw,Dusling:2007gi} \nocite{Peschanski:2009tg} 
as the method of choice for simulating the bulk dynamics of the QGP.  

In viscous hydrodynamics the energy-momentum tensor in the local rest frame is naturally anisotropic with 
${\cal P}_T \neq {\cal P}_L$, where ${\cal P}_T$ and ${\cal P}_L$ are the local rest frame transverse 
and longitudinal pressures, respectively.  The relative amount of momentum-space anisotropy is encoded in the 
shear tensor and in viscous hydrodynamics one has the freedom to choose, in addition to initial temperature and flow 
profiles, an initial value for the shear profile.  In the last years it has emerged that phenomenological predictions 
are rather insensitive to the assumed initial shear profile, see e.g. \cite{Shen:2011eg}.  Further studies have 
shown that the period of large momentum-space anisotropy can persist for a few fm/c \cite{Ryblewski:2012rr}.
In addition, it can be shown that momentum-space anisotropies will be particularly large at the longitudinal 
and transverse edges of the plasma where the system is dilute and poorly approximated by 
na\"ive viscous hydrodynamical treatments \cite{Martinez:2009mf}.

In order to properly address the question of the evolution of systems that possess large momentum-space
anisotropies, one needs to go beyond traditional viscous hydrodynamical treatments.  Viscous hydrodynamics 
relies on an implicit assumption that the shear correction is small and that one can linearize around the isotropic 
ideal background.  If the shear correction is large, a new framework is needed.  This has motivated the 
development of reorganizations of viscous hydrodynamics in which one incorporates the possibility of large 
momentum-space anisotropies into the leading order of the approximation \cite{Florkowski:2010cf,Martinez:2010sc,%
Ryblewski:2010bs,Martinez:2010sd,Ryblewski:2011aq,Martinez:2012tu,Ryblewski:2012rr}.   The framework developed 
has been dubbed {\em anisotropic hydrodynamics}.  This method is capable of describing ideal hydrodynamics through 
free streaming in a single framework in which the expansion is organized around the smallness of the off-diagonal 
components of the energy-momentum tensor.

\section{Formalism}

If we have a system that is azimuthally symmetric in local rest frame (LRF) momenta, then the
energy momentum tensor can be expressed in terms a timelike four-vector $u^\mu$ and a
spacelike four-vector $z^\mu$ which are mutually orthogonal \cite{Ryblewski:2011aq,Martinez:2012tu}
\begin{equation}
T^{\mu\nu} = ({\cal E} +{\cal P}_T) u^\mu u^\nu  - {\cal P}_T g^{\mu \nu}
+ ({\cal P}_L - {\cal P}_T) z^\mu z^\nu \, ,
\label{eq:speroidaltmunu}
\end{equation}
where ${\cal E}$ is the energy density, ${\cal P}_T$ is the transverse pressure, and ${\cal P}_L$
is the longitudinal pressure.  The spacelike four-vector $z^\mu$ is directed along the beamline direction
of a heavy ion collision and $u^\mu$ is the four-velocity of the LRF.

One can derive dynamical equations for the energy density, pressures, and $u^\mu$ by taking moments of the 
Boltzmann equation $p^\mu \partial_\mu f(x,p) = - C[f]$ where $f$ is the one-particle distribution
function and $C$ is the collision kernel.  The moments are defined by multiplying the left and right hand 
sides of the Boltzmann equation by various powers of the four-momentum and then integrating over 
momentum space.  This can be achieved via the $n^{\rm th}$ moment integral operator ${\hat{\cal I}}_n  
\, = \int d\chi \, p^{\mu_1} p^{\mu_2} \cdots p^{\mu_n}$ where $n\geq 0$ is an integer and 
\begin{equation}
\int \! d\chi = \int \! \! \frac{d^4{\bf p}}{(2\pi)^3} \, \delta(p_\mu p^\mu - m^2) \, 2 \theta(p^0) 
= \int \! \! \frac{d^3{\bf p}}{(2\pi)^3} \frac{1}{p^0} \, .
\end{equation}
The resulting zeroth moment of the Boltzmann equation can be compactly written as
\begin{equation}
D n + n \theta = J_0 \, ,
\label{eq:zerothmomgen}
\end{equation}
where $D = u^\mu \partial_\mu$, $\theta = \partial_\mu u^\mu$, $n$ is the number density, 
and $J_0 = (2\pi)^{-3} \int d^3{\bf p} \, p^\mu f / p_0$ is a particle number source.  This equation 
governs the evolution of the number density.  In number conserving theories $J_0=0$; however, in 
number non-conserving theories such as QCD, $J_0 \neq 0$.

The first moment of the Boltzmann equation is equivalent to the requirement of energy and momentum
conservation
\begin{equation}
\partial_\mu T^{\mu\nu} = 0 \, ,
\end{equation}
Taking projections parallel and tranverse to $u^\mu$ one obtains
\begin{equation}
D {\cal E} + ({\cal E} +{\cal P}_\perp) \theta + ({\cal P}_L - {\cal P}_\perp)  u_\nu D_L z^\nu = 0 \, ,
\label{eq:speroidaleq1}
\end{equation}
and
\begin{eqnarray}
&&
({\cal E} +{\cal P}_\perp) D u^\alpha - {\nabla}^\alpha {\cal P}_\perp 
 + z^\alpha D_L ({\cal P}_L - {\cal P}_\perp) + z^\alpha ({\cal P}_L - {\cal P}_\perp) \theta_L
 \nonumber \\
&&
\hspace{1.5cm}
+ ({\cal P}_L - {\cal P}_\perp)  D_L z^\alpha 
- ({\cal P}_L - {\cal P}_\perp)  u^\alpha u_\nu D_L z^\nu
= 0 \, ,
\end{eqnarray}
respectively, where $D_L = z^\mu \partial_\mu$ and $\theta_L = z^\mu \partial_\mu$. 

In order to proceed one can use symmetries to restrict the form of the one-particle distribution 
function.  In the case that the system is azimuthally symmetric in LRF momenta
and consists of massless particles, it suffices to introduce a single scale, $\Lambda$, and a dimensionless
anisotropy parameter $\xi$ \cite{Martinez:2012tu}.   These parameters define a deformation of an arbitrary isotropic
LRF distribution function originally introduced in Ref.~\cite{Romatschke:2003ms}
\begin{equation}
f(t,{\bf x},{\bf p}) = f_{\rm iso}\!\left((p_T^2 + [1 + \xi(t,{\bf x})] p_L^2)/\Lambda^2(t,{\bf x})\right) ,
\label{eq:rsform}
\end{equation}
where we have explicitly indicated that $\Lambda$ and $\xi$ are functions of space and time.  With
this form for the one-particle LRF distribution function and the general equations for the components
of the energy momentum tensor obtained by taking moments of the Boltzmann equation, one can
derive dynamical equations for $\Lambda$ and $\xi$ using 
\begin{equation}
n = \int \frac{d^3{\bf p}}{(2 \pi)^3} f  = \frac{n_{\rm iso}(\Lambda)}{\sqrt{1+\xi}} \, ,
\label{eq:nrs}
\end{equation}
where $n_{\rm iso}$ is the isotropic ($\xi=0$) number density.
One can also evaluate the energy-momentum tensor in the LRF 
\begin{equation}
T^{\mu\nu}= \int \frac{d^3{\bf p}}{(2 \pi)^3} \frac{p^\mu p^\nu}{p_0} f(t,{\bf x},{\bf p}) \, .
\end{equation}
and obtain~\cite{Martinez:2009ry}
\begin{eqnarray}
\label{eq:energyaniso}
{\cal E}(\Lambda,\xi) &=& T^{\tau\tau} = {\cal R}(\xi)\,{\cal E}_{\rm iso}(\Lambda)\, ,\\
\label{eq:transpressaniso}
{\cal P}_T(\Lambda,\xi) &=& \frac{1}{2}\left( T^{xx} + T^{yy}\right) = {\cal R}_T(\xi){\cal P}_{\rm iso}(\Lambda)\, , \\
\label{eq:longpressaniso}
{\cal P}_L(\Lambda,\xi) &=& - T^{\varsigma}_\varsigma = {\cal R}_{\rm L}(\xi){\cal P}_{\rm iso}(\Lambda)\, ,
\end{eqnarray}
where ${\cal P}_{\rm iso}$ and ${\cal E}_{\rm iso}$ are the isotropic ($\xi=0$)
pressure and energy density, respectively, and  
\begin{eqnarray}
{\cal R}(\xi) &=& \frac{1}{2}\left(\frac{1}{1+\xi} +\frac{\arctan\sqrt{\xi}}{\sqrt{\xi}} \right) \, , \\
{\cal R}_T(\xi)  &=& \frac{3}{2 \xi} \left( \frac{1+(\xi^2-1){\cal R}(\xi)}{\xi + 1}\right) \, , \\
{\cal R}_L(\xi) &=& \frac{3}{\xi} \left( \frac{(\xi+1){\cal R}(\xi)-1}{\xi+1}\right) \, .
\end{eqnarray}
The equation of state can be imposed as a relationship between ${\cal E}_{\rm iso}$ and ${\cal P}_{\rm iso}$. 

In the case that the system is boost invariant along the beamline direction and one assumes an ideal equation
of state, the form (\ref{eq:rsform}) results in
\begin{equation}
\frac{1}{1+\xi}D\xi - 6D(\log\Lambda) - 2 \theta = 2 \Gamma\left(1 - {\cal R}^{3/4}(\xi) \sqrt{1+\xi}\right) \, ,
\label{eq:zeromom}
\end{equation}
from the zeroth moment of the Boltzmann equation, where we have used the relaxation time approximation for the 
collision kernel with
a relaxation rate $\Gamma = 2{\cal R}^{1/4}(\xi)\Lambda/5\bar\eta$ with $\bar\eta = \eta/S$ 
with $\eta$ being the shear viscosity and $S$ being the entropy density \cite{Martinez:2010sc}.  The first 
moment in the boost-invariant case becomes
\begin{eqnarray}
&& {\cal R}'(\xi) D\xi + 4 {\cal R}(\xi) D(\log\Lambda) = 
\nonumber
\\ &&  \hspace{2cm}
- \left({\cal R}(\xi) + \frac{1}{3} {\cal R}_\perp(\xi)\right) \Delta_\perp
- \left({\cal R}(\xi) + \frac{1}{3} {\cal R}_L(\xi)\right) \frac{u_0}{\tau} \, ,
\nonumber \\
&& \left[3{\cal R}(\xi) + {\cal R}_\perp(\xi)\right] D u_\perp = 
\nonumber \\ && \hspace{2cm}
-u_\perp \left[ {\cal R}_\perp'(\xi) \tilde{D} \xi 
+ 4  {\cal R}_\perp(\xi) \tilde{D} (\log\Lambda) + \frac{u_0}{\tau} ({\cal R}_\perp(\xi)-{\cal R}_L(\xi)) \right] ,
\nonumber \\
&&
u_y^2 \left[3{\cal R}(\xi) + {\cal R}_\perp(\xi)\right] D \left( \frac{u_x}{u_y} \right) = 
{\cal R}_\perp'(\xi) D_\perp\xi 
+ 4 {\cal R}_\perp(\xi) D_\perp(\log\Lambda) \, ,
\label{eq:firstmom}
\end{eqnarray}
from the first moment, where $\Delta_\perp = \partial_\tau u_0 + \nabla_\perp \cdot {\bf u}_\perp$,
$\tilde{D} = u_0 \partial_\tau + \frac{u_0^2}{u_\perp^2} {\bf u}_\perp \cdot \nabla_\perp$,
$D_\perp = \hat{\bf z} \cdot ({\bf u}_\perp \times \nabla_T) = u_x \partial_y - u_y \partial_x$,
${\bf u}_\perp  = (u_x,u_y)$, and $u_0^2 = 1 + u_\perp^2$.  

\begin{figure}[t]
\includegraphics[width=0.32\textwidth]{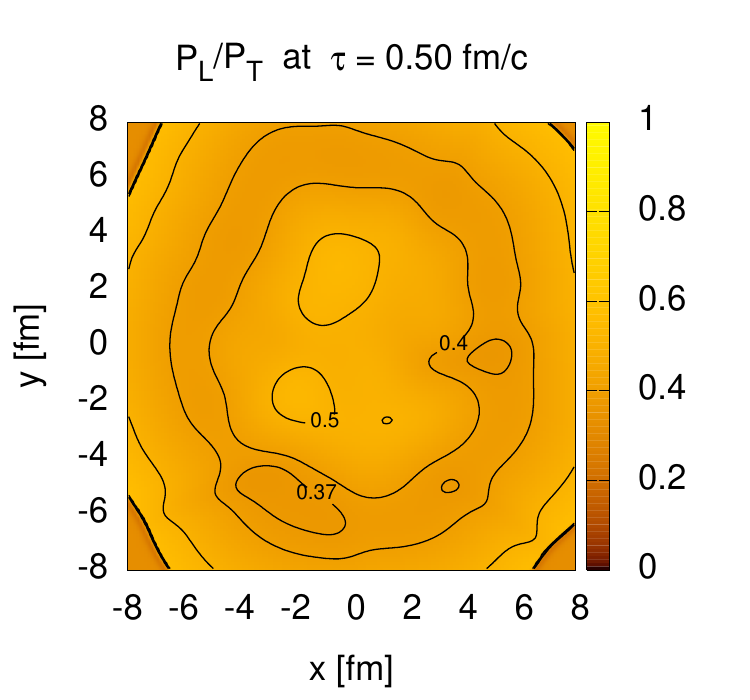}
\includegraphics[width=0.32\textwidth]{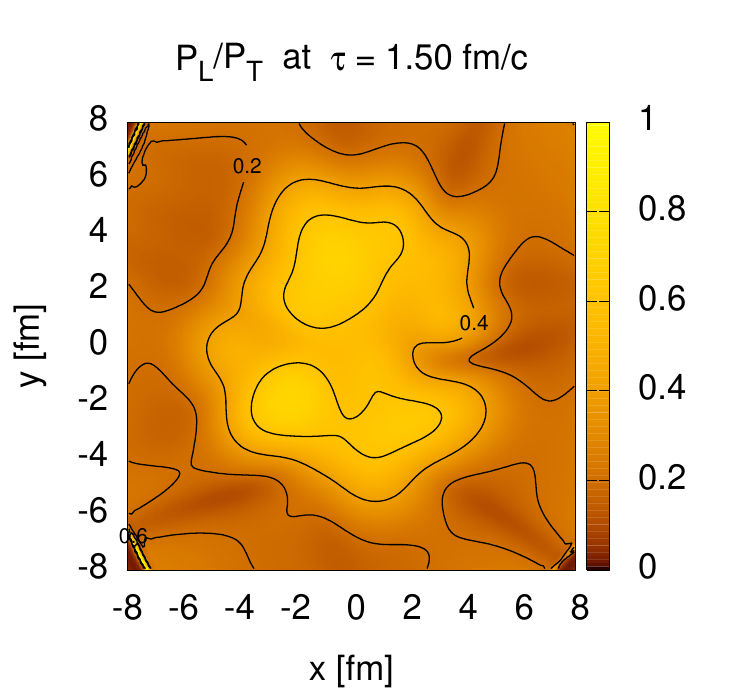}
\includegraphics[width=0.32\textwidth]{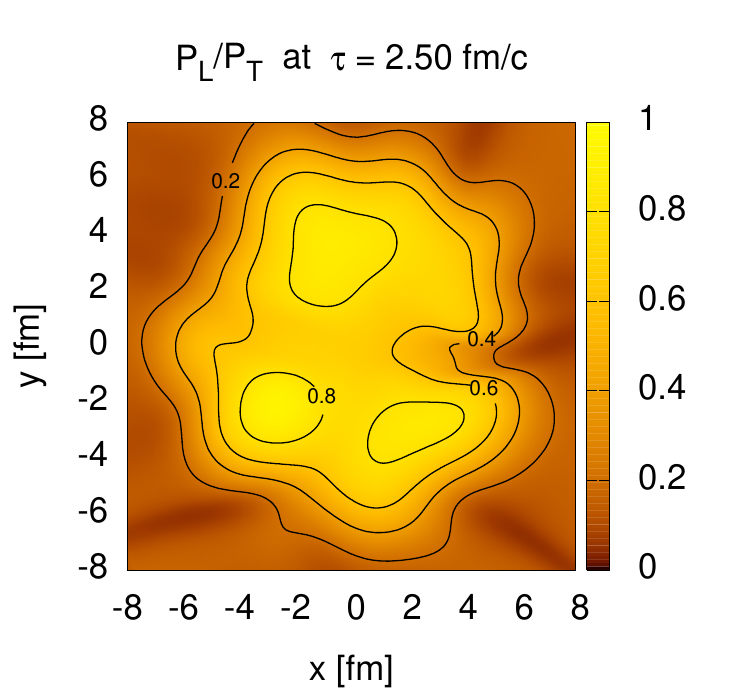}
\caption{(Color online) Evolution of the pressure anisotropy in a (2+1)-dimensional boost-invariant system 
subject to a Monte-Carlo Glauber initial condition.}
\label{fig:f1}
\end{figure}

\section{Results and Outlook}

Having obtained Eqs.~(\ref{eq:zeromom}) and (\ref{eq:firstmom}) one can solve them to find the spatio-temporal
evolution of the energy momentum tensor of an anisotropic system.  By construction the transverse and longitudinal
pressures obtained in the evolution are guaranteed to be positive.  This should be contrasted to solutions to second order
viscous hydrodynamics which can result in negative pressures at early times and near the edges of the matter 
\cite{Martinez:2009mf,Martinez:2010sc}.  In Fig.~\ref{fig:f1} we plot the ratio of the LRF longitudinal and transverse
pressures at three different proper times.  For this figure we assumed a non-central $b=7$ 
fm collision and used a Monte-Carlo sampled Glauber wounded-nucleon profile, a central $b=0$ isotropic temperature 
of $\Lambda_0=T_0=0.6$ GeV at $\tau_0=0.25$ fm/c, $\xi_0 = 0$, and $4\pi\eta/S = 1$.  

As can be seen from Fig.~\ref{fig:f1}, even though the system starts
out being perfectly isotropic at $\tau_0$, after a very short amount of time the system develops a pressure anisotropy
on the order of ${\cal P}_L/{\cal P}_T \sim$ 0.4 -- 0.5 that only slowly relaxes back towards isotropy.  One should
note, importantly, that this figure is generated assuming the best case scenario for isotropization, namely that 
$4\pi\eta/S = 1$.  If one chooses larger values of $\eta$, then one finds larger momentum-space anisotropies at all times.
However, regardless of the value of $\eta$, when evolved using Eqs.~(\ref{eq:zeromom}) and (\ref{eq:firstmom}), the pressures 
(and in particular the longitudinal pressure) remain positive at all times in the entire transverse plane.

In closing I have briefly reviewed the derivation of (2+1)-dimensional anisotropic hydrodynamics.  The full detailed derivation
of the equations along with the numerical algorithms necessary are
contained in Ref.~\cite{Martinez:2012tu}.  The next step in the development of anisotropic hydrodynamics is to 
relax the assumption of azimuthal symmetry of the LRF one-particle distribution function in momentum-space.  This represents
work in progress.

\begin{theacknowledgments}
I thank M. Martinez, R. Ryblewski, and W. Florkowski for collaboration and discussions.  I also thank the organizers 
of the Eleventh Conference on the Intersections of Particle and Nuclear Physics 
(CIPANP 2012). Support for this work was provided by NSF grant No.~PHY-1068765 and the 
Helmholtz International Center for FAIR LOEWE program.
\end{theacknowledgments}

\bibliographystyle{aipproc}   
\bibliography{307_Strickland}

\end{document}